# Relaxometry and dephasing imaging of superparamagnetic magnetite nanoparticles using a single qubit.

*Dominik Schmid-Lorch[†‡], Thomas Häberle[†‡], Friedemann Reinhard[†§], Andrea Zappe[†], Michael Slota[¶], Lapo Bogani[∥], Amit Finkler[†*] and Jörg Wrachtrup[†‡]*

[†]3. Physikalisches Institut, Universität Stuttgart, Pfaffenwaldring 57, 70569 Stuttgart, Germany

[¶]1. Physikalisches Institut, Universität Stuttgart, Pfaffenwaldring 57, 70569 Stuttgart, Germany

[∥]Department of Materials, University of Oxford, 16 Parks Road, OX1 3PH, Oxford, United Kingdom

[‡]Max Planck Institute for Solid State Research, Heisenbergstraße 1, 70569 Stuttgart, Germany

To study the magnetic dynamics of superparamagnetic nanoparticles we use scanning probe relaxometry and dephasing of the nitrogen-vacancy (NV) center in diamond, characterizing the spin-noise of a single 10-nm magnetite particle. Additionally, we show the anisotropy of the NV sensitivity's dependence on the applied decoherence measurement method. By comparing the change in relaxation ($T_1$) and dephasing ($T_2$) time in the NV center when scanning a nanoparticle over it, we are able to extract the nanoparticle's diameter and distance from the NV center using an Ornstein-Uhlenbeck model for the nanoparticle's fluctuations. This scanning-probe technique can be used in the future to characterize different spin label substitutes for both medical applications and basic magnetic nanoparticle behavior.

Magnetic nanomaterials such as magnetic nanoparticles and single molecule magnets (SMM) are of high interest, not only for their applications in basic research, e.g. as interface between classical and quantum physics (SMMs[1]), but also for applications in life sciences. In medicine, chelates of $Gd^{3+}$ ions are already being used as a contrast agent in magnetic resonance imaging (MRI) on a daily basis. Superparamagnetic and magnetic nanoparticles are discussed as an alternative contrast agent and extensive research work is conducted in the field of particle-aided tumor hyperthermia.[2,3]

Hence, the characterization of magnetic behavior on the nanoscale is of major interest. This can be addressed by several established technologies such as superconducting quantum interference device (SQUID) magnetometry. Often these techniques depend on ensemble measurements due to a large detection volume and thus suffer from averaging effects. Novel scanning-probe SQUIDs were demonstrated with loop-diameters as low as 50 nm,[4] giving higher spatial resolution and retaining high magnetic field sensitivity. However, both SQUID magnetometry and other state-of-the-art techniques[5] require cryogenics, which imposes a limitation on the range of observable dynamics.

Magnetic force microscopy (MFM) with a resolution of a few tens of nm and working in ambient conditions is not able to resolve dynamics of the sample, since high frequency (> MHz) components are averaged out by the phase-locked loop (PLL) or phase-detection, a vital part of that measurement technique. In addition, the magnetic back-action on the sample can be rather large thus distorting the results.[6] Fast measurement schemes such as x-ray magnetic circular dichroism (XMCD) can monitor dynamics to a certain degree but are quite challenging from a technical point of view and again are limited in resolution.[7]

Optically read-out single electron spins were proposed as new scanning probe magnetometers to overcome these limitations.[8] Specifically, the nitrogen-vacancy center (NV center) in diamond featuring high sensitivity in a wide frequency range has proven to be a suitable candidate. Imaging was realized from DC magnetic fields[9-11] up to nuclear magnetic spin noise in the MHz regime[12-14] using protocols developed in Ref. 15. Even frequencies in the GHz regime are accessible for imaging, as has been demonstrated in a wide-field setup[16] and recently in a scanning probe approach[17]. In both instances, ensembles of gadolinium ions, which produce wide-band magnetic noise up to 13 GHz, were imaged. In a recent experiment, gadolinium-containing molecules were localized on a diamond substrate and then a single molecule's magnetic noise was shown to inhibit the NV center's relaxation time.[18]

The NV center has also been used to investigate superparamagnetic particles via spectroscopy, most prominently in the form of ferritin molecules[19]. Recently, it was successfully employed to study their temperature dependent dynamics[20].

Here we use a single NV center as a scanning probe to image superparamagnetic magnetite nanoparticles both via relaxometry ($T_1$) and dephasing ($T_2$) contrast. We are able to extract properties such as the particle's diameter and its distance from the NV center, by fitting the measured data with a model that is based on an Ornstein-Uhlenbeck process to describe the dynamics of the particle's magnetization. In addition, we show that relaxometry and dephasing microscopy not only measures magnetic field fluctuations at different frequencies but also along different spatial directions.

The experimental setup is based on a commercial atomic force microscope (AFM) working under ambient conditions and a bulk (100) diamond containing shallow NV centers approximately 5 nm below the surface as shown in Fig. 1a. In all measurements a constant magnetic field of 13 mT was applied, aligned in the direction of the NV axis. The sample is attached to the AFM cantilever, and scanned in contact mode over the diamond surface above an individual NV center. This arrangement is conceptually equivalent[14] to an NV center being fixed to a scanning probe tip as is commonly used for imaging of DC magnetic fields[9,21]. Here, the sample is made of superparamagnetic magnetite nanoparticles (diameter: 8±3 nm, see Supp. Inf.) diluted in sodium silicate, which serves both as a matrix to firmly attach the particles to the cantilever and also as a spacer to keep them separated from each other. To spatially image the particles' characteristic magnetic field fluctuations we use the fact that they induce additional decoherence to the NV center's ground state spin polarization $\vec{S}$.[16,22,23] This results in a decrease of the longitudinal ($T_1$-relaxometry) and transversal ($T_2$-dephasing) spin lifetimes. The NV spin polarization can be analyzed by using the $S_z$ projection (here parallel to the NV center's symmetry axis) and its direct correlation with the fluorescence intensity of the NV center.

Using the $T_1$ relaxation laser pulse sequence illustrated in Fig. 1b, the polarization is initialized and analyzed along the NV axis. This longitudinal spin polarization is sensitive to magnetic field fluctuations $B_\perp$ that are perpendicular to the NV axis and have a frequency component at the NV's Larmor frequency. With the spin echo (SE) sequence for tracking the $T_2$ dephasing, shown in Fig. 1c, the polarization is flipped into the transversal plane via a $\pi/2$ microwave pulse matching the NV's Larmor frequency. The SE measurement method is mostly sensitive to low frequency magnetic field fluctuations along the NV axis $B_\parallel$ only.

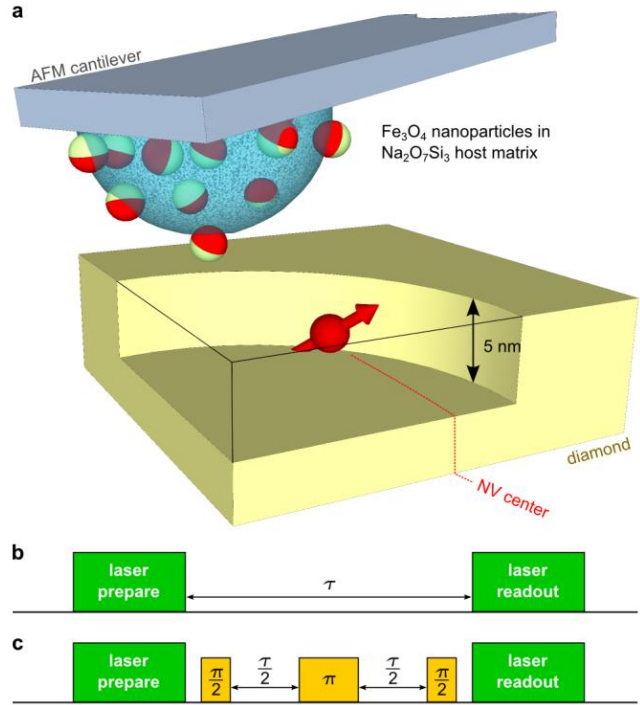

**Figure 1.** Experimental configuration: **a**, Magnetite nanoparticles (green and red spheres) embedded into sodium silicate are attached to an AFM cantilever and scanned across a shallow nitrogen vacancy (NV) center in diamond (red spin). At a distance of less than some tens of nanometers, the local magnetic field fluctuations of the superparamagnetic particles reduce the coherence time of the NV center significantly. Hence, the particles can be imaged using both $T_1$ relaxometry and $T_2$ dephasing contrast. **b,c,** Laser (green) and resonant microwave (yellow) pulse sequences applied to the NV center used for probing. **b**, pulse sequence for $T_1$ relaxation measurement and **c**, pulse sequence for $T_2$ dephasing or spin echo (SE) measurement.

To demonstrate the magnetic field fluctuations from the magnetite nanoparticles full relaxation and dephasing decays (see Fig. 2) were recorded: With a nanoparticle on the AFM cantilever positioned in close proximity to the NV center (red data) and with the sample retracted from the surface by some 10 µm (blue data). The apparent change in $T_1$ and $T_2$ times is confirmed and quantified by fitting the data to theory (see Supp. Inf.). For the $T_1$ time (Fig. 2a) a change by two orders of magnitude from 2344 ± 382 µs to 31 ± 6 µs is obtained when retracting the sample. To evaluate the dephasing time in Fig. 2b one has to consider the echo revival in the data for measurements with retracted sample that is caused by nearby $^{13}C$ nuclear spins (see Supp. Inf.). Hence, we fitted the data with a combined function made up from an envelope and a periodic Gaussian[24]. The resulting $T_2$ time of the NV center varies by a factor of 30, from 0.49 ± 0.05 µs with the sample engaged to 15.16 ± 0.33 µs with the sample retracted.

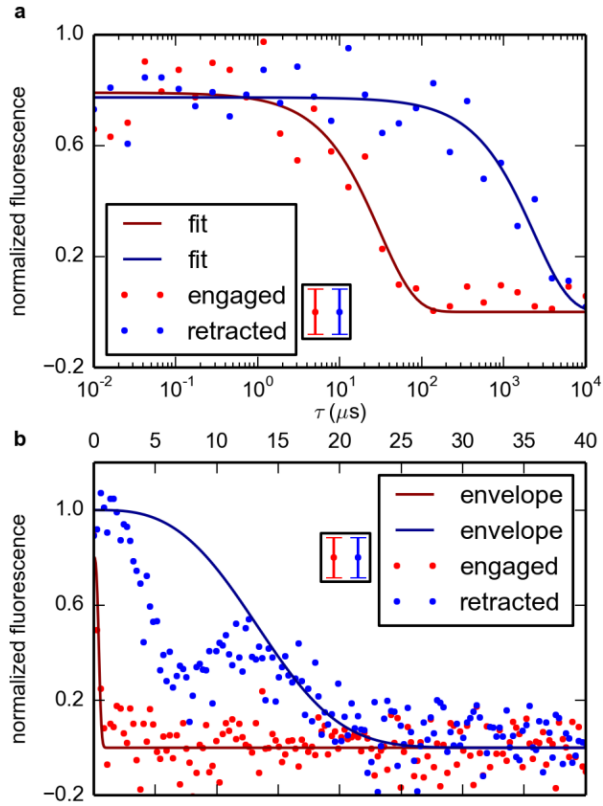

**Figure 2.** Relaxation/dephasing spectroscopy: The NV-center's fluorescence with the magnetite nanoparticle being engaged (red dots) and retracted (blue) from the diamond surface for **a,** $T_1$ relaxation spectroscopy and **b,** $T_2$ dephasing spectroscopy. The solid lines in **a,** are exponential decays fitted to the data. In **b,** the solid lines are envelopes of an extended fitting function taking into account the revival in the measured data. Note the different abscissa time spacing. For clarity, the error bars are displayed next to the legend in each figure.

The distinct changes in $T_1$ and $T_2$ times enable imaging of both relaxation ($T_1$) and dephasing ($T_2$) induced by a single magnetite nanoparticle. This was achieved by recording the relative contrast of the fluorescence intensity (see Supp. Inf.) as a function of the sample position while holding the wait time $\tau$ in the pulse sequences fixed. Figures 3a-c show $T_1$ relaxometry images (500x500 nm with 10 nm pixel size) for three different waiting times $\tau$ and Figs. 3d,e show the associated $T_2$ dephasing for two different waiting times. All five images were acquired during one single AFM scan by cycling through the list of relevant pulse sequences and tagging the respective fluorescence count rates to the corresponding data sets.

Comparing relaxometry with dephasing images, a slight deviation of the nanoparticle image from circular symmetry is visible. We observe a small elongation horizontally in the relaxation ($T_1$) image, whereas its orientation is vertical for the dephasing ($T_2$) measurement method. This is due to the anisotropic nature of the measurement methods detecting either the transverse ($B_\perp$) or the longitudinal fluctuations ($B_\parallel$) with respect to the NV axis. As the NV axis is tilted with respect to the (100) surface orientation this translates into slight asymmetries of the particles image. From magnetic field alignment measurements (see Supp. Inf.) we know that the projection of the NV axis onto the scanning plane is oriented vertically in the images of Fig.3. This is in full agreement with the above observations since for $T_1$ relaxometry the NV center is more sensitive to magnetic field noise in the direction perpendicular to the NV axis. Correspondingly, $T_2$ images are more influenced by noise components in the direction along the NV axis.

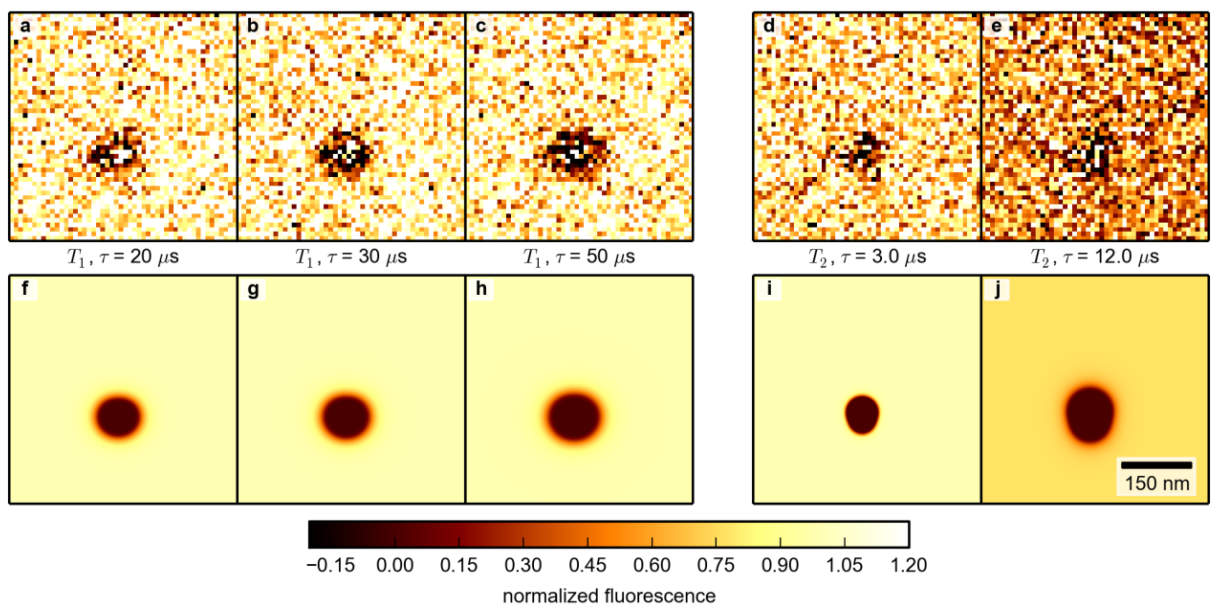

**Figure 3.** Relaxometry and dephasing imaging of a superparamagnetic nanoparticle: **a-e,** Scans across a magnetite particle. Depending on the measurement method, change in fluorescence contrast corresponds to change in **a-c,** spin population or **d,e,** spin coherence. **a-c,** $T_1$ relaxometry at three different wait times τ and **d,e,** $T_2$ dephasing at two different τ. **f-j,** Simulated fluorescence contrast fitted to the experimental data above. All images have the same length scale that is given in **j** and have the same color encoding. Comparing $T_1$ with $T_2$ ones reveals a 90 degrees flipped elongation of images caused by the anisotropic transfer functions of the applied measurement methods. The bright pixels in the center of the dark area are caused by data scaling (see Supp. Inf.).

For modelling, we assume a spherical, superparamagnetic particle of arbitrary size, material and position relative to the NV center (see Supp. Inf. for particle size characterization). To yield maps of $T_1$ and $T_2$ times as a function of the nanoparticle's position, we modeled the magnetic field fluctuations and fitted a simulation of the resulting NV center decoherence to the acquired images. Instead of calculating decoherence for a large number of τ values, this allows us to use few τ values and yet obtain the full decoherence map. The calculation reveals further insight into characteristics of our sample, e.g. the diameter of the particles.

For shallow implanted NV centers the respective decoherence rate, $\Gamma_{tot} = 1/T$, can be divided into two parts and written as

$$\Gamma_{tot} = \Gamma_{int} + \Gamma_{ext} \, . \tag{1}$$

The intrinsic decay rate $\Gamma_{int}$ is attributed to decoherence sources in the diamond lattice such as $^{13}$C isotopes and is considered constant for a given NV center. The external part $\Gamma_{ext}$ accounts for decoherence due to the fluctuating magnetic field of the superparamagnetic nanoparticles close to the NV center and can be calculated via Ref. 20

$$\Gamma_{ext} = \gamma^2 \langle B^2 \rangle \cdot \int S(\nu, T, E) \cdot F_i(\nu) d\nu \, . \tag{2}$$

Here, $\gamma$ is the gyromagnetic ratio of the NV center, $\langle B^2 \rangle$ the second moment of the magnetic field, $S(\nu, T, E)$ the power spectral density and $F_i(\nu)$ describes the filter function of the applied microwave pulse scheme, with $\nu$ being the frequency, $T$ the temperature and $E$ the anisotropy energy barrier.

The superparamagnetic particle at room-temperature is modeled as a sphere made up of magnetic moments $\vec{\mu}_i$, pointing in the same direction, which can be translated into a material dependent magnetic moment density $\rho_{\vec{\mu}}$. Assuming a net magnetic moment of 4 $\mu_B$ per formula unit (Fe$_3$O$_4$) as is typically calculated for magnetite and taking the density of magnetite to be 5.175 g/cm$^3$, this magnetic moment density turns out to be $5.4 \cdot 10^{28}$ $\mu_B/m^3$.[25] Integration over the particle's volume $V$ yields the total magnetic field

$$\vec{B} = \rho_{\vec{\mu}} \frac{\mu_0}{4\pi} \frac{V}{d^3} (-\sin\theta_\mu \cos\phi_\mu, -\cos\theta_\mu \sin\phi_\mu, -2\cos\theta_\mu), \tag{3}$$

depending on the distance $d$ from the particle and the orientation of the particles magnetization given by $(\theta_\mu, \phi_\mu)$. The applied measurement method defines whether the measurement is sensitive to noise parallel to the NV axis $B_\parallel$ (T$_2$ measurement) or perpendicular to the NV axis $B_\perp$ (T$_1$ measurement). Integration over all possible orientations of the particle's magnetization finally yields the second moment:

$$\langle B_\parallel^2 \rangle = \frac{1}{3}\rho_{\vec{\mu}}^2 \frac{\mu_0^2}{16\pi^2} \frac{V^2}{d^6} \left(1 + 3\cos^2(\theta_{NV,\vec{d}})\right) \tag{4}$$

$$\langle B_\perp^2 \rangle = \frac{1}{3}\rho_{\vec{\mu}}^2 \frac{\mu_0^2}{16\pi^2} \frac{V^2}{d^6} \left(5 - 3\cos^2(\theta_{NV,\vec{d}})\right) \tag{5}$$

with $\vec{d}$ being the vector between the NV center and the position of the particle, $\theta_{NV,\vec{d}}$ the angle between NV axis and $\vec{d}$.

The dynamics of the particle's magnetization can be described as an Ornstein-Uhlenbeck process[26] giving the power spectral density,

$$S(\nu, T, E) = \frac{2}{\pi} \frac{\tau_N(T,E)}{1 + \tau_N^2(T,E)\nu^2} \tag{6}$$

governed by the Néel relaxation time $\tau_N(T,E)$[27,28] with the anisotropy energy barrier $E = KV$:

$$\tau_N = \tau_0 \exp\frac{KV}{k_B T} \tag{7}$$

Here $V$ again is the particle's volume and $k_B T$ the thermal energy. The inverse attempt frequency $\tau_0$ is not very well known and ranges from $10^{-9}$ to $10^{-13}$ s[29,30]. The anisotropy constant $K$ for bulk material also has a quite wide range, e.g. $10-41$ kJ m$^{-3}$ for magnetite[2,31].

The filter function $F_i(\nu)$ is defined by the applied measurement method, such as T$_1$ relaxometry[16] or spin echo (SE) spectroscopy[22]. The filter function for T$_1$ relaxometry is given by:[20]

$$F_1(\nu) = \frac{1}{\pi} \frac{1/T_2^*}{(1/T_2^*)^2 + (\nu - \nu_{+1})^2} + \frac{1}{\pi} \frac{1/T_2^*}{(1/T_2^*)^2 + (\nu - \nu_{-1})^2} . \tag{8}$$

$T_2^*$ is the NV dephasing rate under a Ramsey experiment and $\nu_{\pm 1}$ are the microwave transition frequencies for $m_s = 0 \to m_s = \pm 1$. The SE filter function $F_2(\nu)$ is dominated by the wait-time between microwave pulses $\tau/2$:

$$F_2(\nu) = \frac{1}{4\pi^2} \frac{2}{\tau} \frac{\sin^4\left(2\pi\frac{\nu}{4}\tau\right)}{\left(\frac{\nu}{4}\right)^2} . \tag{9}$$

The combination of formulae (1)-(9) allows for the simulation of the NV center's decoherence rate under the influence of a nanoparticle.

Next, we fitted this simulation to the obtained fluorescence contrast images (Figs. 3a-e) with the free parameters minimum vertical distance, $z$, between the particle and the NV center and the particle radius, $r$, to match all five images at once. Values for the inverse attempt frequency $\tau_0 = 1.0 \cdot 10^{-13}$ s and the anisotropy constant $K = 26$ kJ/m$^3$ were chosen from the wide range of literature values by conducting a short study on their influence on the fit (see Supp. Inf.). The simulated images are shown in Figs. 3f-j. The resulting fit parameters are $z = 16.3 \pm 4.7$ nm and $r = 7.06 \pm 0.4$ nm corresponding to $59576 \pm 10128$ iron atoms. Additionally, the relaxation times of the intrinsic decoherence are T$_{1,\text{int}} = 2713 \pm 311$ µs and T$_{2,\text{int}} = 19 \pm 1$ µs as derived from the NV center without influence of the nanoparticles magnetic noise.

In the simulated images the slight elongation due to the anisotropy in the transfer function is seen clearly. From measuring the NV axis orientation by rotating a stationary external field we know its orientation to be as shown in Fig. 3. Small elongations of the $T_2$ images (Figs. 3i,j) in vertical direction are influenced by this. There is a larger spread due to the higher sensitivity to fluctuations from magnetic field fluctuations in vertical direction. Additionally, the anisotropy in the transfer function of the $T_1$-spot can be quantitatively evaluated by its elliptical eccentricity $\varepsilon$. The simulation gives a value of $\varepsilon = 0.4$ for the skewed NV-axis in the present (100)-diamond. In the hypothetical case of the NV axis pointing exactly towards the surface, like in a (111)-diamond, this value would read $\varepsilon = 0.0$ and for the case of the NV axis being parallel to the surface $\varepsilon = 0.5$ (see Supp. Inf.). From our measurements we get the eccentricity $\varepsilon = 0.44 \pm 0.14$. The large error is due to the shot noise in the data acquisition.

Figures 4a,b show the simulated relaxation ($T_1$) and dephasing ($T_2$) times of the NV center related to each position in the measured images (Figs. 3a-e). To further substantiate that the simulation quantitatively describes the decoherence, we measured the $T_1$ and $T_2$ times along a line scan in this map with the protocol used for acquiring Fig. 2. The cantilever was held stationary while the recording was done at each sample position. The relaxation times obtained via fits to the data are plotted in Figs. 4c,d together with the values of the simulation (green line) and show good agreement.

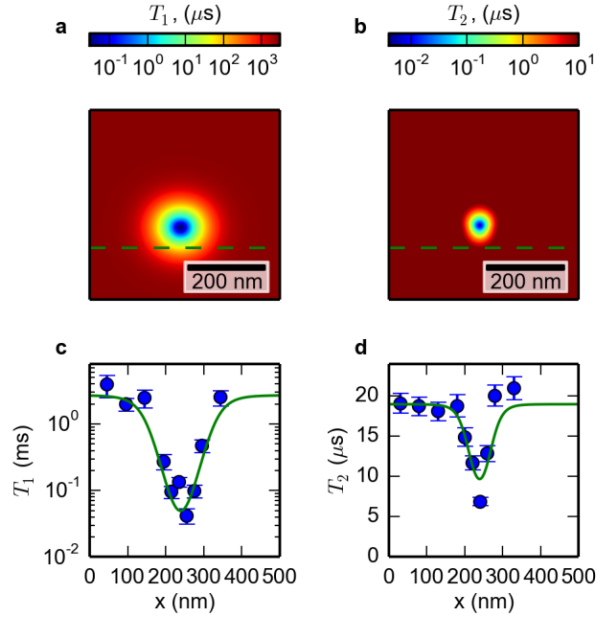

**Figure 4.** Relaxation and dephasing time map: **a,b**, Decoherence time maps from simulations of the superparamagnetic nanoparticle fitted to the scanned images Figs. 3a-e for **a**, $T_1$ relaxation and **b**, $T_2$ dephasing. **c,d,** Measured relaxation, **c,** and dephasing, **d,** times based on evaluations of full time spectra (like in Fig. 2) at discrete positions along a line scan. The solid green lines give the simulated time values from the above maps (values taken along dashed lines in **a,** and **b,**). We also estimated the maximal values for the second moments of the magnetic field, $\langle B_\perp^2 \rangle$ and $\langle B_\parallel^2 \rangle$, that were measured within the dynamic range of the $T_1$ and $T_2$ measurements for the extracted distance between the NV center and the nanoparticle. These values were calculated excluding the artifact points where $T_1$ and $T_2$ are too short to quantify. The values are: $\langle B_\perp^2 \rangle = 8.8 \ mT^2$ and $\langle B_\parallel^2 \rangle = 1.8 \ mT^2$.

In summary, we have demonstrated combined nanoscale relaxometry and dephasing imaging of magnetite nanoparticles under ambient conditions. We probe the magnetic particles' magnetic field fluctuation at two different frequency ranges while scanning the sample, which provides firm basis for dynamic simulations. The anisotropy of the measurement of magnetic field noise observed in $T_1$ and $T_2$ maps agrees with the NV center's axis orientation with respect to the experimental setup's principal scanning directions.

Acquiring relaxation images as in the present case is time consuming since it is mainly limited by the photon flux from the single NV center and the limited contrast of the spin signal. For future experiments it is promising to further improve acquisition by a single shot NV readout[32], thereby shortening the measurement time up to 3 orders of magnitude[33]. This would reduce acquisition time to some tens of milliseconds per pixel. The signal-to-noise ratio could be further increased by optimizing the collection efficiency of the fluorescence light with diamond nanopillars containing shallow NVs[34]. An interesting application of this imaging technique could be nanoscale imaging of living cells containing ferritin as a natural non-toxic contrast agent featuring similar magnetic field fluctuations[19] and thus omitting the side effects of gadolinium ions. Additionally, different spin label agents could be compared on single particle basis for their use as alternative MRI contrast agents or their application in particle aided tumor hyperthermia. Especially the ability to perform the characterization at ambient conditions, which closely resemble the conditions at which the contrast agents are expected to work as such, makes it a viable and efficient way of screening different spin-labels for medicinal purposes – other methods cannot perform these dynamics studies at these conditions. For nanoparticle research this investigation technique could assist in imposing stricter upper and lower bounds on the big range of literature values for the attempt frequency and the anisotropy constant.

ASSOCIATED CONTENT

**Supporting Information**. Detailed description of the experimental setup and the data acquisition. Overview on sample and $T_2$ dephasing fitting function. Dependence of the transfer function on the NV axis orientation. SQUID susceptometry measurements. Study of the fit parameters. Measurements of the collapses and revivals of the $^{13}C$ nuclei at different fields. Significance of simultaneous acquisition of $T_1$ and $T_2$ for the fit.

AUTHOR INFORMATION

**Corresponding Author**

*E-mail: a.finkler@physik.uni-stuttgart.de

**Present Addresses**

§Walter Schottky Institut, E24, Technische Universität München, Am Coulombwall 4, D-85748 München, Germany

**Author Contributions**

‡These authors contributed equally to this work.

F.R. and J.W. conceived the experiment and with A.F. supervised the project. D.S.-L. and T.H. conducted the experiments. D.S.-L. and A.Z. prepared the samples. T.H. performed the simulations. T.H. and A.F. analyzed the data. M.S. performed the SQUID susceptibility measurements and L.B. analyzed them. D.S.-L., T.H., F.R., L.B., A.F. and J.W. wrote the manuscript.

ACKNOWLEDGMENT

The authors acknowledge support from EU via ERC grant SQUTEC and integrated projects DIADEMS and SIQS, DARPA (QuASAR), the DFG via research group 1493, SFB/TR21 and SPP 1601 and contract research of the Baden-Württemberg foundation. The authors thank K. Karrai for the helpful discussions on operating the attoCSFM device; and J. Greiner for the theoretical and mathematical support. The authors also thank I. Platzman from the Max Planck Institute for Intelligent Systems for his assistance in acquiring SEM images. A.F. acknowledges support from the Alexander von Humboldt Foundation.

ABBREVIATIONS

NV, nitrogen vacancy; AFM, atomic force microscope; SMM, single molecule magnet; MRI, magnetic resonance imaging; SQUID, superconducting quantum interference device; MFM, magnetic force microscopy; XMCD, x-ray magnetic circular dichroism; SE, spin echo.